\documentclass[10pt,journal,compsoc]{IEEEtran}
\usepackage{amsmath,amsfonts}
\usepackage{algorithmic}
\usepackage{algorithm}
\usepackage{array}
\usepackage[caption=false,font=normalsize,labelfont=sf,textfont=sf]{subfig}
\usepackage{textcomp}
\usepackage{stfloats}
\usepackage{url}
\usepackage{verbatim}
\usepackage{graphicx}
\usepackage{cite}
\usepackage{soul}
\usepackage[colorlinks=true]{hyperref}
\usepackage{threeparttable/threeparttable}
\usepackage{xcolor}
\newcommand{\methodname}{{\fontfamily{ppl}\selectfont
SimSID}}
\newcommand{\oldmethodname}{{\fontfamily{ppl}\selectfont
SQUID}}
\definecolor{Highlight}{HTML}{000000}  

\newenvironment{jlrev}{\color{Highlight}}{}
\newcommand{\eg}{\mbox{e.g.,\ }}
\newcommand{\etal}{\mbox{et al.}}
\newcommand{\ie}{\mbox{i.e.,\ }}
\newcolumntype{P}[1]{>{\centering\arraybackslash}p{#1}}
\newlength\savewidth\newcommand\shline{\noalign{\global\savewidth\arrayrulewidth
  \global\arrayrulewidth 0.8pt}\hline\noalign{\global\arrayrulewidth\savewidth}}

\begin{document}

\title{Exploiting Structural Consistency of Chest Anatomy for Unsupervised Anomaly Detection\\in Radiography Images}

\author{Tiange Xiang*, 
        Yixiao Zhang*, 
        Yongyi Lu,
        Alan Yuille,~\IEEEmembership{Senior~Member,~IEEE,} 
        Chaoyi Zhang, \\
        Weidong Cai,~\IEEEmembership{Member,~IEEE,}
        Zongwei Zhou,~\IEEEmembership{Member,~IEEE}
\IEEEcompsocitemizethanks{
\IEEEcompsocthanksitem *Tiange Xiang and Yixiao Zhang contribute equally.
\IEEEcompsocthanksitem Corresponding author: Zongwei Zhou (\href{mailto:zzhou82@jh.edu}{\textsc{zzhou82@jh.edu}}).
\IEEEcompsocthanksitem T. Xiang, C. Zhang, W. Cai are with the School of Computer Science, University of Sydney, Camperdown NSW 2006, Australia.\protect\\
\{txia7609@uni,czha5168,tom.cai\}.sydney.edu.au; lchen025@e.ntu.edu.sg
\IEEEcompsocthanksitem Y.~Zhang, Y.~Lu, A.~Yuille, Z.~Zhou are with the Department of Computer Science, Johns Hopkins University, Baltimore, MD 21218 USA.\protect\\
\{yzhan334,zzhou82\}@jh.edu; \{yylu1989,alan.l.yuille\}@gmail.com}
}

\markboth{IEEE TRANSACTIONS ON PATTERN ANALYSIS AND MACHINE INTELLIGENCE, 2024}%
{Shell \MakeLowercase{\textit{et al.}}: Bare Demo of IEEEtran.cls for Computer Society Journals}

\IEEEtitleabstractindextext{%
\begin{abstract}
Radiography imaging protocols focus on particular body regions, therefore producing images of great similarity and yielding recurrent anatomical structures across patients.
Exploiting this structured information could potentially ease the detection of anomalies from radiography images. 
To this end, we propose a \ul{Sim}ple \ul{S}pace-Aware Memory Matrix for \ul{I}n-painting and \ul{D}etecting anomalies from radiography images (abbreviated as \methodname). We formulate anomaly detection as an image reconstruction task, consisting of a space-aware memory matrix and an in-painting block in the feature space.
During the training, \methodname\ can taxonomize the ingrained anatomical structures into recurrent visual patterns, and in the inference, it can identify anomalies (unseen/modified visual patterns) from the test image.
Our \methodname\ surpasses the state of the arts in unsupervised anomaly detection by +8.0\%, +5.0\%, and +9.9\% AUC scores on ZhangLab, COVIDx, and CheXpert benchmark datasets, respectively.
Code: \href{https://github.com/MrGiovanni/SimSID}{https://github.com/MrGiovanni/SimSID} 

\end{abstract}


}

\maketitle

\section{Introduction} \label{sec:introduction}

\begin{figure}
    \centering
    \includegraphics[width=1\linewidth]{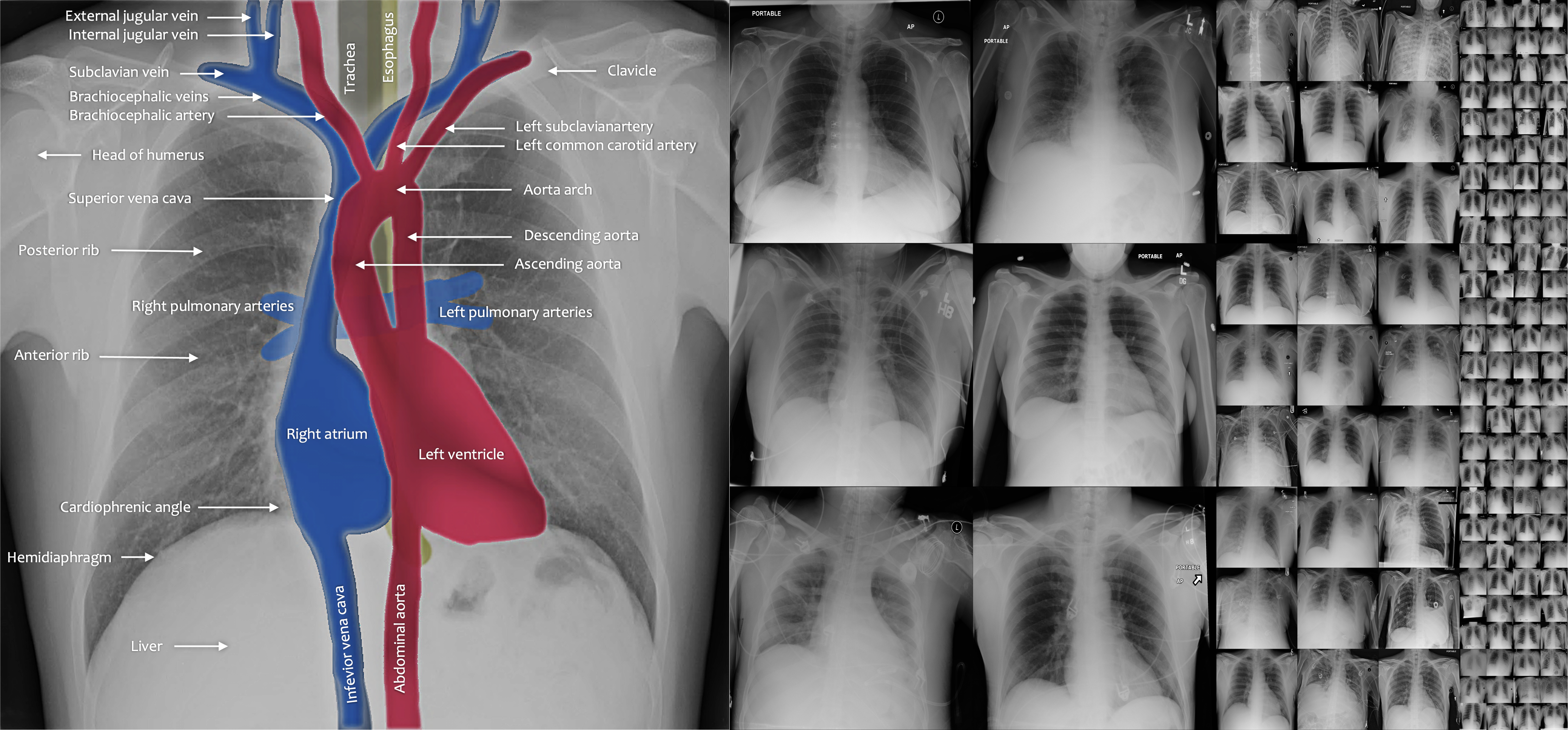}
    \caption{
    Anomaly detection in radiography images can be both easier and harder than photographic images.
    It is easier because radiography images are spatially structured due to consistent imaging protocols.
    It is harder because anomalies are subtle and require medical expertise to annotate.
    We contribute a novel anomaly detection method (\methodname) that directly exploits the structured information in radiography images.
    }
    \label{fig:introductory_figure}
\end{figure}

Vision tasks in photographic and radiographic images differ significantly.  In photographic object identification, the object's location within the image is typically less important---a cat remains a cat regardless of its position within the image. Conversely, in radiography, the relative location and orientation of anatomical structures are crucial for both identifying normal anatomy and recognizing pathologies~\cite{zhao20213d,haghighi2021transferable,haghighi2020learning,zhou2022interpreting,zhou2021towards}. 
Due to standardized imaging protocols in radiography, images exhibit a high degree of similarity across patients, equipment manufacturers, and institutions (see examples in \figureautorefname~\ref{fig:introductory_figure}). 
Consistent and recurrent anatomy can facilitate the analysis of numerous critical problems and should be considered a significant advantage of radiography imaging.
For example, several investigations have demonstrated the value of harnessing this prior knowledge to enhance Deep Nets' performance, such as adding location features, modifying objective functions, and constraining coordinates relative to landmarks in images~\cite{smoger2015statistical,anas2016automatic,mirikharaji2018star,zhou2019integrating,lu2020learning,feng2020parts2whole,zhou2019models}. 
This paper focuses on unsupervised anomaly detection, seeking to answer the critical question: 
\emph{Can we exploit consistent anatomical patterns and their spatial information to strengthen Deep Nets in detecting anomalies from radiography images without manual annotation?}

Unsupervised anomaly detection only uses healthy images for model training and requires no other annotations such as disease diagnosis or localization~\cite{baur2021autoencoders}.
As many as 80\% of clinical errors occur when the radiologist misses the abnormality in the first place \cite{brady2017error}. The clinical impact of anomaly detection is to reduce that 80\% by clearly pointing out to radiologists that there exists a suspicious lesion and then having them look at the scan in depth. 
We formulate the task of anomaly detection as an in-painting task to exploit the anatomical consistency in appearance, position, and layout in the chest region. 
Specifically, we propose a \ul{Sim}ple \ul{S}pace-Aware Memory Matrix for \ul{I}n-painting and \ul{D}etecting anomalies from radiography images (abbreviated as \methodname).
In the training phase, our model can \emph{dynamically} maintain a visual pattern dictionary by taxonomizing the recurrent anatomical structure based on its spatial locations. 
Due to the consistency in anatomy, the same body region across normal images is expected to express similar visual patterns, which makes the total number of unique patterns manageable.
In the inference, since anomaly patterns are not present in the learned dictionary, the reconstructed image is expected to be unrealistic.
As a result, the model can identify the anomaly by assessing the quality of the in-painting task.
The success of anomaly detection has two basic assumptions~\cite{zimek2017outlier}: \emph{first}, \textcolor{Highlight}{anomalies only occur very rarely in the training data} {\jlrev (or a small proportion in an image)};
\emph{second}, anomalies differ from the normal patterns significantly.
Consequently, the learned dictionary will reflect the general distribution of anatomical patterns in normal human anatomy.
Notably, our \methodname\ is robust to a level of abnormal images in the training set by automatically omitting minority anatomical patterns (evidenced in \figureautorefname~\ref{fig:ratio_robustness}).
This should be considered a significant advantage because it can largely relax the requirement of disease-free images for training existing unsupervised anomaly detection methods~\cite{baur2021autoencoders}.

We have conducted extensive experiments on \emph{three} large-scale radiography imaging datasets.
Our \methodname\ is significantly superior to 21 predominant methods in unsupervised anomaly detection by at least 8.0\% on the ZhangLab dataset, yielding an AUC of 91.1\%;
5.0\% AUC gain on the COVIDx dataset, with an AUC of 83.5\%;
additionally, we have demonstrated a 9.9\% improvement over the state of the arts on the Stanford CheXpert dataset, with an AUC of 79.7\%.
The quantitative results and qualitative visualization show the superiority of \methodname\ over the state of the arts.

\section{Related Work} \label{sec:related_work}

\subsection{Anomaly Detection in Natural Imaging}

Anomaly detection is the task of identifying rare events that deviate from the distribution of normal data~\cite{omar2013machine}. 
Early attempts include one-class SVM~\cite{scholkopf1999support}, dictionary learning~\cite{zhao2011online}, and sparse coding~\cite{cong2011sparse}.
Due to the lack of sufficient samples of anomalies, later works typically formulate anomaly detection as an unsupervised learning problem~\cite{li2013anomaly,ruff2018deep,zong2018deep,sidibe2017anomaly,hendrycks2016baseline, lee2017training,liang2017enhancing,lee2018simple,devries2018learning,hendrycks2018deep}. These can be roughly categorized into reconstruction-based and density-based methods. Reconstruction-based methods train a model (\eg Auto-Encoder) to recover the original inputs~\cite{chen2018unsupervised,siddiquee2019learning,tang2021disentangled,zhou2021models,zavrtanik2021draem,ristea2021self}. The anomalies are identified by subtracting the reconstructed image from the input image. Density-based methods predict anomalies by estimating the normal data distribution (\eg via VAEs~\cite{kingma2013auto} or GANs~\cite{akcay2018ganomaly,schlegl2019f}). However, their learned distribution for normal images cannot explain possible abnormalities. In this paper, we address these limitations by maintaining a visual pattern dictionary which is extracted from homogeneous medical images. 

Several other previous works investigated the use of image in-painting for anomaly detection, \ie parts of the input image are masked out and the model is
trained to recover the missing parts in a self-supervised way~\cite{reiss2021panda,nguyen2021unsupervised,zavrtanik2021reconstruction, haselmann2018anomaly,sato2022anatomy}. 
There are also plenty of works on detecting anomalies in video sequences~\cite{hasan2016learning, lu2019future, liu2021hybrid, acsintoae2021ubnormal}. Recently, Bergmann~\etal~\cite{bergmann2020uninformed} and Salehi~\etal~\cite{salehi2021multiresolution} proposed student-teacher networks similar to ours, whereas our method utilizes such a structure to distillate input-aware features only, and the teacher network is completely disabled during inference. 

\subsection{Anomaly Detection in Medical Imaging}

Anomaly detection in the medical domain is usually approached at per pathology-basis~\cite{pang2022editorial,hu2023label,li2023early,liu2023clip,zhang2023continual}. There are \textit{supervised} anomaly detection methods to identify specific types of abnormalities, such as  vascular lesions~\cite{zuluaga2011learning}, malignant melanoma~\cite{khan2021attributes}, brain tumors~\cite{bakas2018identifying,zhou2019unet++}, and pulmonary nodules/embolism~\cite{zheng2019automatic,islam2021seeking}. Recent \textit{unsupervised} anomaly detection methods have been proposed to detect anomalies in general~\cite{fernando2020deep,tschuchnig2021anomaly,baur2021autoencoders,heer2021ood,xiao2022delving}.
With the help of GANs, anomaly detection can be achieved with \textit{weak} annotation. In AnoGAN~\cite{schlegl2017unsupervised}, the discriminator was heavily over-fitted to the normal image distribution to detect the anomaly. Subsequently, f-AnoGAN~\cite{schlegl2019f} was proposed to improve computational efficiency. Naval~\etal~\cite{naval2021implicit} designed an autoencoder network to fit the distribution of normal images. The spatial coordinates and anomaly probabilities are mapped over a proxy for different tissue types. Han~\etal~\cite{han2021madgan} proposed a two-step GAN-based framework for detecting anomalies in MRI slices as well. However, their method relies on a voxel-wise representation for the 3D MRI sequences, which is impossible in our task. Most recently, a hybrid framework SALAD~\cite{zhao2021anomaly} was proposed that combines GAN with self-supervised techniques. Normal images are first augmented to carry the forged anomaly through pixel corruption and pixel shuffling. The fake abnormal images, along with the original normal ones, are fed to the GAN for learning more robust feature representations. However, these approaches demand strong prior knowledge and assumptions about the anomaly type to make the augmentation effective. 

Incorporating memory modules into neural networks has been demonstrated to be effective for many tasks ~\cite{kumar2016ask, fan2019heterogeneous, kaiser2017learning, cai2018memory, lee2018memory}. Adopting a Memory Matrix for unsupervised anomaly detection was first proposed in MemAE~\cite{gong2019memorizing}. In addition to auto-encoding (AE), Gong~\etal\ injected an extra Memory Matrix between the encoder and the decoder to capture normal feature patterns during training. The matrix is jointly optimized along with the AE and hence learns an essential basis to be able to assemble normal patterns. Based on this paradigm, Park~\etal~\cite{park2020learning} introduced a non-learnable memory module that can be updated with inputs.
Considering the extra memory usage in existing methods, Lv~\etal~\cite{Lv2021MPN} proposed a dynamic prototype unit that encodes normal dynamics on the fly, while consuming little additional memory.

\textcolor{Highlight}{With the recent progress in diffusion models~\cite{kazerouni2022diffusion,linmans2024diffusion,yang2023diffusion,du2023boosting}, it is also feasible to achieve anomaly detection by generating normal image samples. One of the earliest attempts that followed this paradigm for medical anomaly detection was proposed by Wolleb \etal \cite{wolleb2022diffusion}. SynDiff \cite{ozbey2023unsupervised}, as one of the following-up methods, extended the generation quality further by incorporating adversarial projections during the inverse diffusion process.}

Differing from photographic images, radiography imaging protocols produce images with consistent anatomical patterns, and meanwhile, the anomalies in radiography images can be subtle in appearance and hard to interpret (\figureautorefname~\ref{fig:introductory_figure}). 
Unlike most existing works, we present a novel method that explicitly harnesses the radiography images' properties, therefore dramatically improving the performance in anomaly detection from radiography images.

\begin{figure*}[!t]
    \centering
    \includegraphics[width=\linewidth]{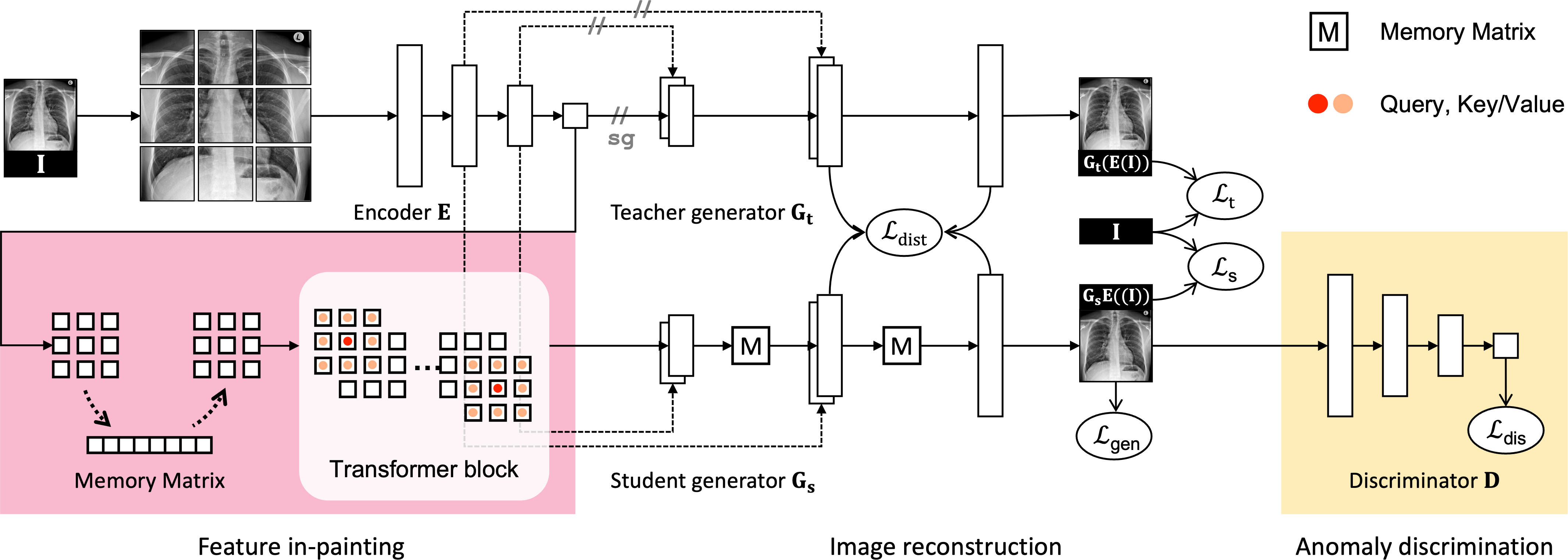}
    \caption{\textbf{\methodname \ overview.} We divide an input image into $N\times N$ non-overlapping patches and feed them into the encoder for feature extraction. 
    Two generators will be trained to reconstruct the original image. 
    Along with the reconstruction, a dictionary of anatomical patterns will be created and updated dynamically via a novel space-aware memory matrix (\S\ref{sec:queue});
    The teacher generator directly uses the features extracted by the encoder;
    the student generator uses the features augmented by a new feature in-painting block (\S\ref{sec:inpaint}). 
    The teacher and student generators are coupled through a knowledge distillation paradigm.
    We employ a discriminator to assess whether the image reconstructed by the student generator is real or fake.
    Once trained, the discriminator can be used to detect anomalies in test images (\S\ref{sec:alert}).
    }
    \label{fig:framework}
\end{figure*}

\subsection{Our Previous Work}

We first presented \ul{S}pace-aware Memory \ul{Qu}eues for \ul{I}n-painting and \ul{D}etecting anomalies from radiography images (\oldmethodname) published in CVPR-2023~\cite{xiang2023squid}. This paper is a significant extension with the following four improvements.
\begin{enumerate}
    \item We have introduced new notations, formulas, and diagrams, as well as detailed methodology descriptions along with their learning objectives, for a succinct framework overview.
    \item We have significantly simplified the framework by removing the Memory Queue and masked shortcut while achieving higher performance and easing the training than SQUID~\cite{xiang2023squid}.
    \item We have examined \methodname\ with 21 existing unsupervised (and also one weakly-supervised) anomaly detection methods on three radiography imaging tasks, showing that \methodname\ surpasses all these methods by a large margin, as well as \oldmethodname~\cite{xiang2023squid}.
    \item We have investigated the robustness of our \methodname\ to the normal/abnormal ratio in the training set, relaxing the requirement of the disease-free training set of existing anomaly detection approaches.
\end{enumerate}

\section{\methodname}
\label{sec:method}

\subsection{Overview} \label{sec:framework}

\noindent\textbf{Feature extraction:} 
We divide the input image into $N\times N$ non-overlapping patches, then use a CNN encoder to extract features for each patch. 
The extracted features will be used for image reconstruction. 
Practically, the encoder can be any backbone architectures, and for simplicity, we adopt basic Convolutions and Pooling layers in the experiments. 

\medskip\noindent\textbf{Feature in-painting:}
A dictionary of normal anatomical patterns will be created and updated dynamically through a Memory Matrix (\S\ref{sec:queue}). The extracted patch features will be substituted by the most close items in the matrix. Then, the substituted features of each image patch would be masked out, a transformer block (\S\ref{sec:inpaint}) is used to predict the masked feature based on the surrounding patch features.

\medskip\noindent\textbf{Image reconstruction:}
We introduce teacher and student generators to reconstruct the original image. 
Specifically, the teacher generator reconstructs the image using the features extracted by the encoder directly (essentially an auto-encoder~\cite{rumelhart1985learning}). 
The student generator, on the other hand, reconstructs the image using the features augmented by our in-painting block. 
The teacher and student generators are coupled through knowledge distillation~\cite{hinton2015distilling} at all the up-sampling levels.
The objective of the student generator is to reconstruct a normal image from the augmented features; the reconstructed image will then be used for anomaly discrimination (\S\ref{sec:alert});
while the teacher generator\footnote{We disabled the backpropagation between the teacher generator and encoder by stop-gradient~\cite{he2020momentum} and showed its benefit in \tableautorefname~\ref{tab:component}.} serves as a regularizer to prevent the student generator from collapsing\footnote{Alternative strategies to avoid collapse include early stopping, elastic regularization, or Gaussian prior~\cite{reiss2021panda,anas2016automatic}.} (constantly generating the same normal image). 

\medskip\noindent\textbf{Anomaly discrimination:}
Following the adversarial learning~\cite{schlegl2017unsupervised, schlegl2019f}, we employ a discriminator to assess whether the generated image is real or fake.
Both teacher and student generators will receive the gradient derived from the discriminator.
The two generators and the discriminator are competing against each other in a way that, together, they converge to an equilibrium.
Once trained, the discriminator can be used to detect anomalies in test images (\S\ref{sec:alert}).

\subsection{Developing Space-aware and Hierarchical Memory}
\label{sec:queue}

\noindent\textbf{Motivation:} 
The Memory Matrix was initially introduced by Gong~\etal~\cite{gong2019memorizing} and has since been widely adopted in unsupervised anomaly detection~\cite{liu2021hybrid,zaheer2021anomaly,gong2021memory,zhou2021memorizing,kim2021robust}.
To forge a ``normal'' appearance, the features are \textit{augmented} by weighted averaging the similar patterns in Memory Matrix.
This augmentation is, however, applied to the features extracted from the whole image, discarding the location and spatial information embedded in images.
Therefore, Memory Matrix in its current form cannot perceive the anatomical consistency that radiography images can offer.

\begin{figure}[!t]
    \centering
    \includegraphics[width=\linewidth]{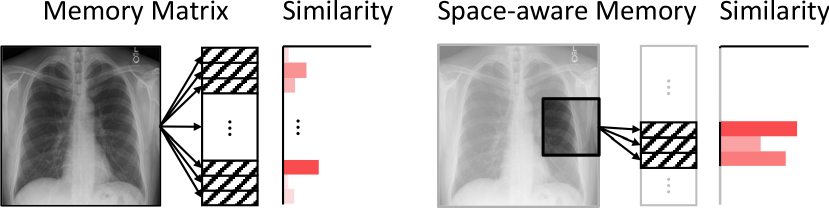}
    \caption{\textbf{Space-aware memory.} For unique encoding of location information, we restrict each patch to be only able to access a set of specific tokens in the memory.
    }
    \label{fig:space_mem}
\end{figure}

\medskip\noindent\textbf{Space-aware memory:} 
To harness the spatial information, we pass the divided patches into the model rather than the entire image.
These patches are associated with unique location information of the original image.
We seek to build the relationship between the patch location and memory region. The memory matrix $\mathbf{M}$ is divided into blocks $\{\mathbf{M}_{i,j} \in \mathbb{R}^{N \times C}\}$, each associated with a patch at location $(i, j)$, where $N$ and $C$ denote the number and the dimension of items, respectively. Let $\mathbf{z}_{i,j} \in \mathbb{R}^C$ denote the feature of patch $(i,j)$, we obtain the augmented feature $\mathbf{\hat{z}}$ as follows:
\begin{equation}
    \mathbf{\hat{z}_{i,j}} = \sum_{k=1}^{N} G(s^k)\mathbf{M}_{i,j}^k,
\end{equation}
where $s^k$ is the similarity score computed by dot product between the patch feature $\mathbf{z}_{i,j}$ and the $k$-th memory item $\mathbf{M}_{i,j}^k$. $G(\cdot)$ is the Gumbel-softmax operation, which shrinks the number of activated memory items\footnote{Controlling the number of activated memory items has proven to be advantageous for anomaly detection~\cite{graves2014neural}. 
However, setting a hard shrinkage threshold as in \cite{gong2019memorizing} fails to adapt to cases where abnormal signals are sufficient to reconstruct a normal image.
Inspired by \cite{jang2016categorical}, we present a \textit{Gumbel Shrinkage} schema: only activating the top-$k$ most similar memory items during the forward pass and distributing the gradient to all patterns during back-propagation.
Gumbel Shrinkage improves AUC from 86.2\% to 91.1\% (see \tableautorefname~\ref{tab:component}).}.
With the division of image patches together with memory blocks, a patch derived from a particular location can only search for similar items within a specific block in the Memory Matrix (illustrated in~\figureautorefname~\ref{fig:space_mem}).
We refer to this new searching strategy as ``space-aware memory''.
This strategy can also accelerate the searching speed compared with~\cite{gong2019memorizing} as it no longer has to go through the entire Memory Matrix to assemble similar features.
Results in \tableautorefname~\ref{tab:component} highlight the significance of space-aware memory (AUC improved from 77.6\% to 91.1\%).

\begin{figure}[!t]
    \centering
    \includegraphics[width=0.85\linewidth]{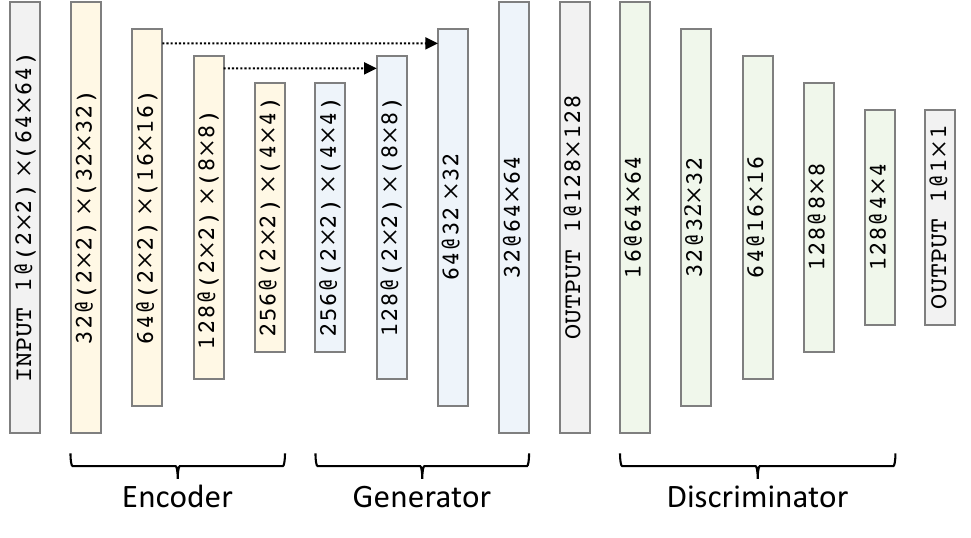}
    \caption{\textbf{\methodname \ architecture.} Our \methodname\ consists of an encoder, a student generator, a teacher generator, and a discriminator. All of the network architectures are built with plain convolution, batch normalization, and ReLU activation layers. Given an input image, we first divide it into non-overlapping patches. The encoder then extracts the patch features.
    The student and teacher generators were constructed identically. The only difference is that additional Memory Matrices are placed in the student generator. 
    The discriminator was constructed in a more lightweight style. Note that the images are discriminated at their full resolution rather than in patches.
    }
    \label{fig:architecture_details}
\end{figure}

\medskip\noindent\textbf{Hierarchical memory:} The use of one memory matrix at the deepest layer in the encoder is insufficient to reconstruct high-quality image with details. 
To capture anatomical patterns at different scales, we placed a space-aware memory matrix at several levels of the generator to create a hierarchy of scales. 
Studies in~\cite{gong2019memorizing} discovered that too many memories can lead to excessive information filtering and degrade the model's capacity to retain the most representative normal patterns instead of all needed ones. 
This problem is solved by adding skip connections between the encoder and the generator\footnote{It is worth noting that the outermost skip connection should not be added (shown in \figureautorefname~\ref{fig:architecture_details}). It is because a memory matrix must be followed by skip connections; otherwise, the reconstruction might be fulfilled by the highest-level encoding-decoding information, making all other lower-level encoding, decoding and memory blocks not work.}.
In each generator layer, the feature map is upsampled and concatenated with low-level features carried by the skip connection, then filtered by the following space-aware Memory Matrix. 
We empirically found that a total of three Memory Matrices (one in the feature in-painting block and two in the generator) are sufficient. 
This design is also proved to be effective by~\cite{liu2021hybrid} in flow-guided video anomaly detection.
Results in \tableautorefname~\ref{tab:component} highlight the significance of hierarchical memory and skip connection, achieving an AUC improvement of 8.2\% and 11.6\%.

\subsection{In-painting Features by Learned Memory Matrix}
\label{sec:inpaint}

\noindent\textbf{Motivation:}
Image in-painting~\cite{pathak2016context} was initially proposed to recover corrupted regions in the image based on the available neighboring context.
The recovered regions, however, have been seen to associate with boundary artifacts, distorted and blurry predictions, particularly when using methods based on Deep Nets~\cite{liu2018image,zavrtanik2021reconstruction}.
These undesired artifacts are responsible for numerous false positives when formulating anomaly detection as an image in-painting task~\cite{siddiquee2019learning,zhou2021models}. 
It is because the subtraction between input and output will reveal artifacts generated by Deep Nets instead of true anomalies.
To alleviate this issue, we propose the in-painting task at the feature level rather than the image pixel level.
Latent features are invariant to subtle noise, rotation, and translation in the pixel level and therefore are expected to be more suitable for anomaly detection. The model predicts central features based on neighboring features. This in-painting step is repeated for all of the patch neighborhoods through sliding-window with stride of 1. Similar to the sliding-window as in convolutions, the whole process is fully parallelizable and computed efficiently.

\begin{figure}[!t]
    \centering
    \includegraphics[width=\linewidth]{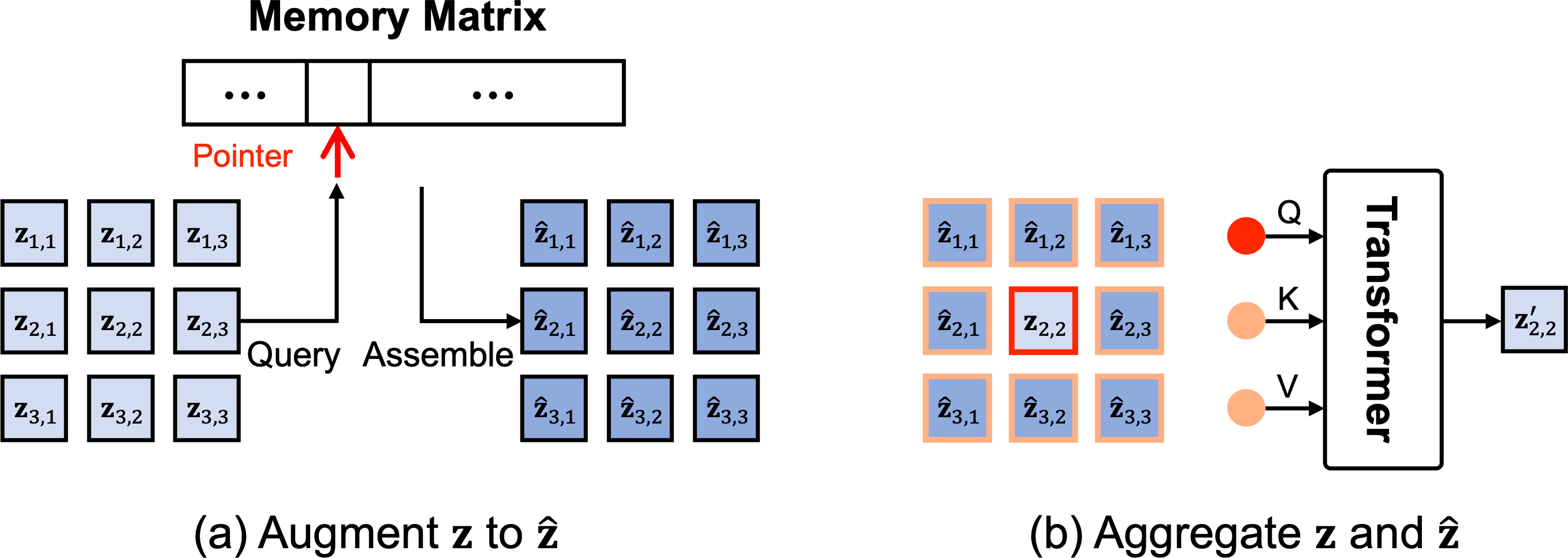}
    \caption{
    \textbf{Two-step workflow of the in-painting block.} (a) Each non-overlapping patch feature $\mathbf{z}$ is queried to an unique region in Memory Matrix, the most similar items are assembled to $\mathbf{\hat{z}}$. (b) Each center patch feature $\mathbf{z}$ and its eight neighbors $\mathbf{\hat{z}}$ are used as query and key/value respectively to a Transformer layer for in-painting. During training, the Memory Matrix is updated through optimization via backpropagation.
    }
    \label{fig:inpaint_block}
\end{figure}

\medskip\noindent\textbf{In-painting block:} 
We integrate our Memory Matrix with a novel in-painting block to perform an in-painting task at the feature level. The $w\times h$ non-overlapping patch features $\mathbf{z}_{\{(1,1),\cdots,(w,h)\}}$ are augmented to the most similar ``normal'' patterns $\mathbf{\hat{z}}_{\{(1,1),\cdots,(w,h)\}}$ in Memory Matrix (\figureautorefname~\ref{fig:inpaint_block}a). 
Since $\mathbf{\hat{z}}$ is assembled by patterns from previously seen images, it is not subject to the current input image. 
To recap characteristics of the current image, naturally, we aggregate both patch features $\mathbf{z}$ and augmented features $\mathbf{\hat{z}}$ using a Transformer block~\cite{vaswani2017attention}. 
For each patch $\mathbf{z}_{i,j}$, its spatially adjacent eight ``normal'' patches $\mathbf{\hat{z}}_{\{(i-1,j-1),\cdots,(i+1,j+1)\}}$ are used as conditions to refine $\mathbf{z}_{i,j}$ (\figureautorefname~\ref{fig:inpaint_block}b).
The query token is flattened $\mathbf{z}_{(i,j)}\in\mathbb{R}^{1\times *}$ and key/value tokens are $\mathbf{\hat{z}}_{\{(i-1,j-1),\cdots,(i+1,j+1)\}}\in\mathbb{R}^{8\times *}$. 
At the start and the end of our in-painting block, we apply an extra pair of point-wise convolutions ($1\times 1$ convolutional kernel) to reduce feature dimensions and accelerate the training process.
\subsection{Anomaly Discrimination} \label{sec:alert}

Our in-painting block focuses on augmenting any patch feature (either normal or abnormal) into the normal feature pattern. 
The student generator will then reconstruct a ``normal'' image based on the augmented features.
The teacher generator is used to preserve the normal image intact and prevent the student generator from collapsing. 
Once trained, the semantic (rather than pixel-level) difference between the input and the reconstructed image is expected to be small if normal; the semantic difference will be big if there are anomalies.
We therefore delegate the optimized discriminator network for alerting anomalies perceptually. \textcolor{Highlight}{Unlike common approaches that use pixel-level comparisons to alert anomaly \cite{zhao2021anomaly}, we are trying to utilize a discriminator to assess the generation of 'normal samples'. This allows \methodname\ to be more robust to pixel-level noise and variations.} For better clarification, we notate the encoder, teacher generator, student generator, and discriminator as $\mathbf{E}$, $\mathbf{G}_{\text{t}}$, $\mathbf{G}_{\text{s}}$, and $\mathbf{D}$. An anomaly score ($A$) can be computed through:
\begin{equation}
    A = \phi(\frac{\mathbf{D}(\mathbf{G}_{\text{s}}(\mathbf{E}(\mathbf{I}))) - \mu}{\sigma}),   
\end{equation}
where $\phi(\cdot)$ is the Sigmoid function, $\mu$ and $\sigma$ are the mean and standard deviation of anomaly scores calculated on all training samples. 

\subsection{Loss Function} \label{sec:loss}

Our \methodname\ is optimized by five loss functions. 
The mean square error (MSE) between input and reconstructed images is used for both teacher and student generators.
Concretely, for the teacher and student generators, we have:
\begin{equation}
    \mathcal{L}_{\text{t}} = ||\mathbf{I} - \mathbf{G}_{\text{t}}(\mathbf{E}(\mathbf{I}))||^2,\quad \mathcal{L}_{\text{s}} = ||\mathbf{I} - \mathbf{G}_{\text{s}}(\mathbf{E}(\mathbf{I}))||^2,
\end{equation}
where $\mathbf{I}$ denotes the input image.
Following the knowledge distillation paradigm, we apply a distance constraint between the teacher and student generators at all levels:
\begin{equation}
    \mathcal{L}_{\text{dist}} = \sum^{l}_{i=1}(\mathbf{z}_{\text{t}}^i - \mathbf{z}_\text{s}^i)^2,
\end{equation}
where $l$ is the level of features used for knowledge distillation, $\mathbf{z}_{\text{t}}$ and $\mathbf{z}_{\text{s}}$ are the intermediate features in the teacher and student generators, respectively.
In addition, we employ an adversarial loss (similar to DCGAN~\cite{radford2015unsupervised}) to improve the quality of the image generated by the student generator.
Specifically, the following equation is minimized: 
\begin{equation}
    \mathcal{L}_{\text{gen}} = \log(1-\mathbf{D}(\mathbf{G}_{\text{s}}(\mathbf{E}(\mathbf{I})))).
\end{equation}
The discriminator seeks to maximize the probability for real images and the inverted probability for fake images: \begin{equation}
    \mathcal{L}_{\text{dis}} = \log(\mathbf{D}(\mathbf{I})) + \log(1-\mathbf{D}(\mathbf{G}_{\text{s}}(\mathbf{E}(\mathbf{I})))).
\end{equation}

In summary, our \methodname\ is trained to \emph{minimize} the generative loss terms ($\lambda_{\text{t}}\mathcal{L}_{\text{t}} + \lambda_{\text{s}}\mathcal{L}_{\text{s}} + \lambda_{\text{dist}}\mathcal{L}_{\text{dist}} + \lambda_{\text{gen}}\mathcal{L}_{\text{gen}}$)
and to \emph{maximize} the discriminative loss term ($\lambda_{\text{dis}}\mathcal{L}_{\text{dis}}$).

\begin{table*}[!t]
\footnotesize
\centering
\caption{Benchmark results on the \textit{official} test sets of the three datasets.
    Apart from those performances directly taken from other literature, we present the mean and standard deviation (mean$\pm$s.d.) across three different trials for all models.
    For every dataset, the AUC improvement between our \methodname\ and the best alternative baseline method is significant at $p=0.05$ level, performed by an independent two sample \textit{t}-test. 
    }
    \begin{tabular}{p{0.37\linewidth}p{0.12\linewidth}P{0.12\linewidth}P{0.12\linewidth}P{0.12\linewidth}P{0.12\linewidth}P{0.12\linewidth}}
        Dataset: \emph{ZhangLab} & Ref~\&~Year & AUC~(\%) & Acc~(\%) & F1~(\%) \\
        \shline
        Auto-encoder$^{\dagger}$ & - & 59.9 & 63.4 & 77.2  \\
        VAE$^{\dagger}$~\cite{kingma2013auto} & Arxiv'13 & 61.8 & 64.0 & 77.4  \\
        Ganomaly$^{\dagger}$~\cite{akcay2018ganomaly} & ACCV'18& 78.0 & 70.0 & 79.0  \\
        f-AnoGAN$^{\dagger}$~\cite{schlegl2019f} & MIA'19 & 75.5 & 74.0 & 81.0 \\
        MemAE~\cite{gong2019memorizing}& ICCV'19 & 77.8$\pm$1.4 & 56.5$\pm$1.1 & 82.6$\pm$0.9 \\
        Fixed-Point GAN$^{\ddagger}$~\cite{siddiquee2019learning} & ICCV'19 & 83.1 & 78.0 & 84.3 \\
        MNAD~\cite{park2020learning} & CVPR'20 & 77.3$\pm$0.9 & 73.6$\pm$0.7 & 79.3$\pm$1.1\\
        SALAD$^{\dagger}$~\cite{zhao2021anomaly} & TMI'21& 82.7$\pm$0.8 & 75.9$\pm$0.9 & 82.1$\pm$0.3  \\
        CutPaste~\cite{li2021cutpaste} & CVPR'21 &  73.6$\pm$3.9 & 64.0$\pm$6.5 & 72.3$\pm$8.9  \\
        PANDA~\cite{reiss2021panda} & CVPR'21 & 65.7$\pm$1.3 & 65.4$\pm$1.9 & 66.3$\pm$1.2  \\
        M-KD~\cite{salehi2021multiresolution} & CVPR'21 & 74.1$\pm$2.6 & 69.1$\pm$0.2 & 62.3$\pm$8.4  \\
        IF 2D~\cite{naval2021implicit} & MICCAI'21 & 81.0$\pm$2.8& 76.4$\pm$0.2& 82.2$\pm$2.7\\
        PaDiM~\cite{defard2021padim} & ICPR'21 & 71.4$\pm$3.4 & 72.9$\pm$2.4 &80.7$\pm$1.2\\
        IGD~\cite{chen2021deep} & AAAI'22 & 73.4$\pm$1.9 & 74.0$\pm$2.2 & 80.9$\pm$1.3\\
        \hline
        SQUID~\cite{xiang2023squid} & Ours (CVPR'23) & 87.6$\pm$1.5 & 80.3$\pm$1.3 & 84.7$\pm$0.8 \\
        \methodname & Ours & \textbf{91.1$\pm$0.9} & \textbf{85.0$\pm$1.0} & \textbf{88.0$\pm$1.1} \\
    \end{tabular}
    \begin{tablenotes}
    \scriptsize
        \item $^{\dagger}$The results are taken from Zhao~\etal~\cite{zhao2021anomaly}; $^{\ddagger}$Fixed-Point GAN is considered as a baseline of weakly supervised learning (requiring image-level labels)
    \end{tablenotes}
    \vspace{1.5em}
    \begin{tabular}{p{0.37\linewidth}p{0.12\linewidth}P{0.12\linewidth}P{0.12\linewidth}P{0.12\linewidth}P{0.12\linewidth}P{0.12\linewidth}}
        Dataset: \emph{COVIDx} & Ref~\&~Year & AUC~(\%) & Acc~(\%) & F1~(\%) \\
        \shline
        DAE$^{*}$~\cite{masci2011stacked} & ICANN'11 & 55.7 \\
        ALAD$^{\dagger}$~\cite{zenati2018adversarially} & ICDM'18 & 58.0 \\
        Ganomaly$^{\dagger}$~\cite{akcay2018ganomaly} & ACCV'18 & 58.4\\
        OCGAN$^{*}$~\cite{perera2019ocgan} & CVPR'18 & 61.2 \\
        f-AnoGAN$^{\ddagger}$~\cite{schlegl2019f}& MIA'19 & 66.9 \\
        MemAE~\cite{gong2019memorizing}& ICCV'19 & 71.8$\pm$3.6 & 77.1$\pm$2.1 & 86.4$\pm$0.8 \\
        ADGAN$^{*}$~\cite{liu2020photoshopping} & ISBI'19 & 65.9 \\
        CCD+IGD$^{*}$~\cite{tian2021constrained} & MICCAI'21 & 74.6 \\
        PaDim$^{*}$~\cite{defard2021padim} & ICPR'21 & 61.4 \\
        PatchCore$^{\dagger}$~\cite{roth2021towards} & Arxiv'21 & 52.0 \\
        CutPaste~\cite{li2021cutpaste} & CVPR'21 & 78.5$\pm$2.3 & \textbf{83.1$\pm$0.4} & \textbf{89.5$\pm$0.2} \\
        PANDA~\cite{reiss2021panda} & CVPR'21 & 72.3$\pm$1.0 & 76.9$\pm$0.8 & 86.4$\pm$0.4 \\
        M-KD~\cite{salehi2021multiresolution} & CVPR'21 & 71.7$\pm$1.1 & 69.7$\pm$4.5 & 55.6$\pm$2.5 \\
        MS-SSIM$^{*}$~\cite{chen2021deep} & AAAI'22 & 63.4 \\
        IGD$^{*}$~\cite{chen2021deep} & AAAI'22 & 69.9 \\
        \hline
        SQUID~\cite{xiang2023squid} & Ours (CVPR'23) & 74.7$\pm$0.9 & 76.8$\pm$0.1 & 86.0$\pm$0.2 \\
        \methodname  & Ours & \textbf{83.5$\pm$0.6} & 82.6$\pm$0.6 & 88.8$\pm$0.1\\
    \end{tabular}
    \begin{tablenotes}
    \scriptsize
        \item $^{*}$The results are taken from Tian~\etal~\cite{tian2022unsupervised}; $^{\dagger}$The results are taken from Rahman Siddiquee~\etal~\cite{siddiquee2021a2b}; $^{\ddagger}$The results are taken from Tian~\etal~\cite{tian2021self}
    \end{tablenotes}
    \vspace{1.5em}
    \begin{tabular}{p{0.37\linewidth}p{0.12\linewidth}P{0.12\linewidth}P{0.12\linewidth}P{0.12\linewidth}P{0.12\linewidth}P{0.12\linewidth}}
        Dataset: \emph{CheXpert} & Ref~\&~Year & AUC~(\%) & Acc~(\%) & F1~(\%) \\
        \shline
        Ganomaly~\cite{akcay2018ganomaly} & ACCV'18 & 68.9$\pm$1.4 & 65.7$\pm$0.2 & 65.1$\pm$1.9\\
        f-AnoGAN~\cite{schlegl2019f}& MIA'19 & 65.8$\pm$3.3 & 63.7$\pm$1.8 & 59.4$\pm$3.8 \\
        MemAE~\cite{gong2019memorizing}& ICCV'19 & 54.3$\pm$4.0 & 55.6$\pm$1.4 & 53.3$\pm$7.0 \\
        CutPaste~\cite{li2021cutpaste} & CVPR'21 & 65.5$\pm$2.2 & 62.7$\pm$2.0 & 60.3$\pm$4.6  \\
        PANDA~\cite{reiss2021panda} & CVPR'21 & 68.6$\pm$0.9 & 66.4$\pm$2.8 & 65.3$\pm$1.5  \\
        M-KD~\cite{salehi2021multiresolution} & CVPR'21 & 69.8$\pm$1.6 & 66.0$\pm$2.5 & 63.6$\pm$5.7  \\
        \hline
        SQUID~\cite{xiang2023squid} & Ours (CVPR'23) & 78.1$\pm$5.1 & 71.9$\pm$3.8 & \textbf{75.9$\pm$5.7} \\
        \methodname & Ours & \textbf{79.7$\pm$2.2} & \textbf{72.9$\pm$1.9} & 71.9$\pm$2.3
    \end{tabular}
    \label{tab:chest_xray_benchmark}
\end{table*}

\section{Experiments}

\subsection{Public Chest Radiography Benchmarks}

\noindent\textbf{ZhangLab Chest X-ray~\cite{kermany2018identifying}:} 
This dataset contains healthy and pneumonia images, \emph{officially} split into training and test sets. 
The training set consists of 1,349 normal and 3,883 abnormal images;
the test set has 234 normal and 390 abnormal images. 
We randomly separate 200 images (100 normal and 100 abnormal) from the training set as the validation set for early-stopping. 
Since the images are of varying sizes, we resized all the images to $128\times128$. We used this dataset for ablation studies as well.

\medskip\noindent\textbf{Stanford CheXpert~\cite{irvin2019chexpert}:} 
We conducted evaluations on the front-view PA images in the CheXpert  dataset, which account for a total of 12 different anomalies. 
In all front-view PA scans, there are 5,249 normal and 23,671 abnormal images for training; 
250 normal and 250 abnormal images (with at least 10 images per disease type) from the training set for testing; 14 normal and 19 abnormal images for early-stopping (\texttt{val} set based on the \textit{official} split). 
All images are resized to $128\times128$ as inputs.

\medskip\noindent\textbf{COVIDx~\cite{wang2020covid}:} The original dataset contains a train and a test set. The train set has 29,187 chest radiographs, of which 8,085 are normal, 5,555 are non-covid pneumonia and 15,547 are COVID-19 positive. The test set has 400 chest X-rays, of which 100 are normal, 100 are non-covid pneumonia and the rest 200 are COVID-19 positive. We randomly separate 400 images (200 normal, 100 non-covid pneumonia and 100 COVID-19 pneumonia) from the training set as the validation set. COVIDx \href{https://github.com/lindawangg/COVID-Net/blob/master/labels/train\_COVIDx9A.txt}{v9} was used in our experiments.

\subsection{Baselines, Metrics, and Implementation}
We considered a total of 21 major baselines for direct comparison (elaborated in~\tableautorefname~\ref{tab:chest_xray_benchmark}): for example, Auto-encoder, VAE~\cite{kingma2013auto}---the classic UAD methods;  Ganomaly~\cite{akcay2018ganomaly}, f-AnoGAN~\cite{schlegl2019f}, IF~\cite{naval2021implicit},
SALAD~\cite{zhao2021anomaly}---the current state of the arts for medical imaging; and MemAE~\cite{gong2019memorizing}, CutPaste~\cite{li2021cutpaste}, M-KD~\cite{salehi2021multiresolution}, PANDA~\cite{reiss2021panda}, PaDiM~\cite{defard2021padim}, IGD~\cite{chen2022deep}---the most recent UAD methods.
We evaluated performance using standard metrics: receiver operating characteristic (ROC) curve, precision-recall (PR) curves, area under the ROC curve (AUC), accuracy (Acc) and F1-score (F1). 
All results were based on at least \emph{three} independent runs.

We utilized common data augmentation strategies such as random translation within the range $[-0.05, +0.05]$ in four directions and a random scaling within the range of $[0.95, 1.05]$. 
The Adam optimizer was used with a batch size of 16 and a weight decay of $1e$-$5$. 
The learning rate was initially set to $1e$-$4$ for both the generator and the discriminator and then decayed to $2e$-$5$ in 200 epochs following the cosine annealing scheduler. 
The discriminator was trained at every iteration, while the generator was trained every two iterations. 
We set the loss weights as $\lambda_{\text{t}}=0.01$, $\lambda_{\text{s}}=10$, $\lambda_{\text{dist}}=0.001$, $\lambda_{\text{gen}}=0.005$, and $\lambda_{\text{dis}}=0.005$.
We divided the input images in $4\times4$ non-overlapping patches generator.
The architectures of our generators and discriminator are detailed in \figureautorefname~\ref{fig:architecture_details}.

\section{Results}

\begin{figure*}[!t]
    \centering
    \includegraphics[width=0.9\linewidth]{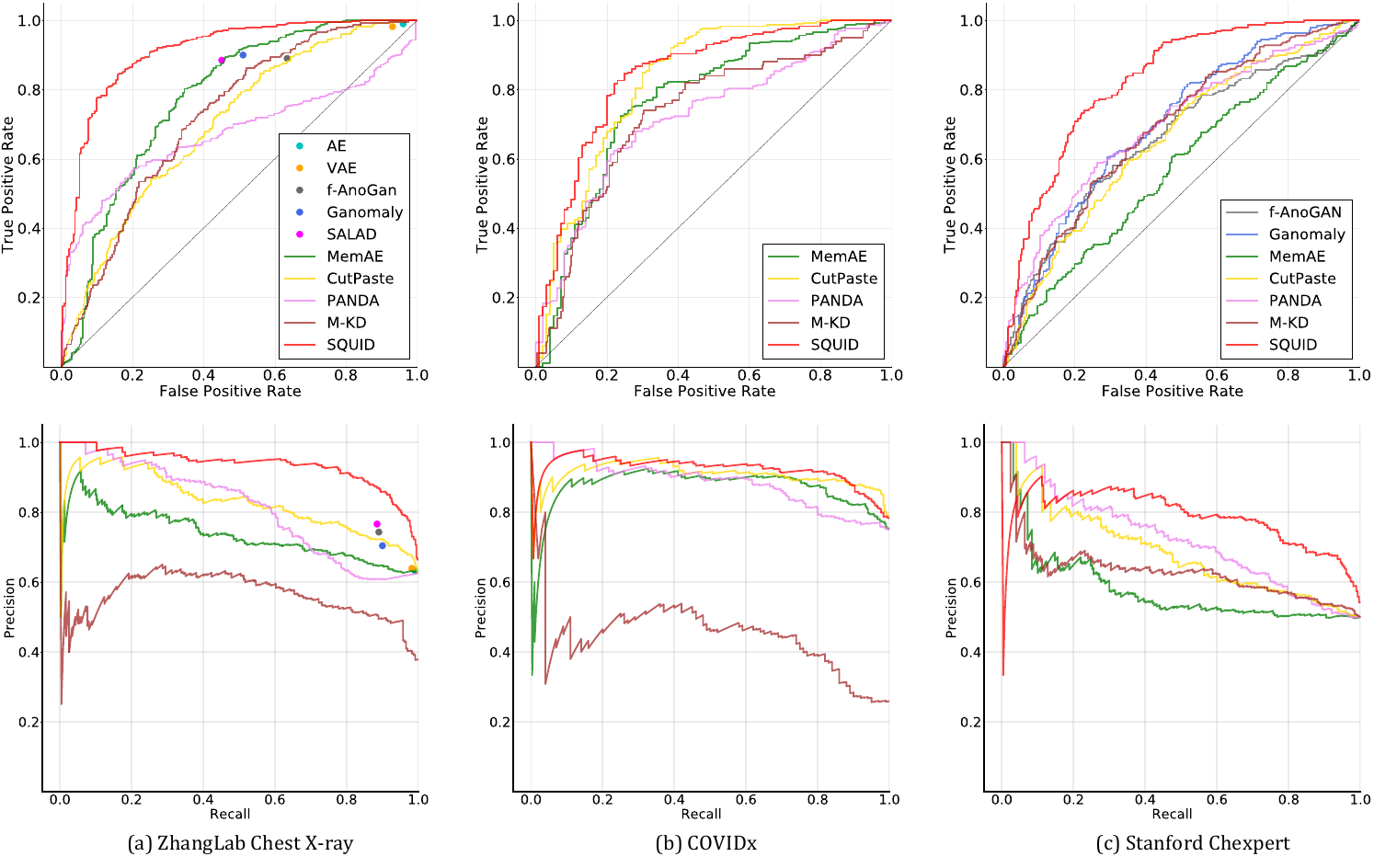}
    \caption{ROC curves and PR space comparison on the ZhangLab Chest X-ray, COVIDx and Stanford CheXpert datasets. ROC = receiver operating characteristic; PR = precision-recall.
    }
    \label{fig:roc}
\end{figure*}

\begin{figure*}[!t]
    \centering
    \includegraphics[width=0.89\linewidth]{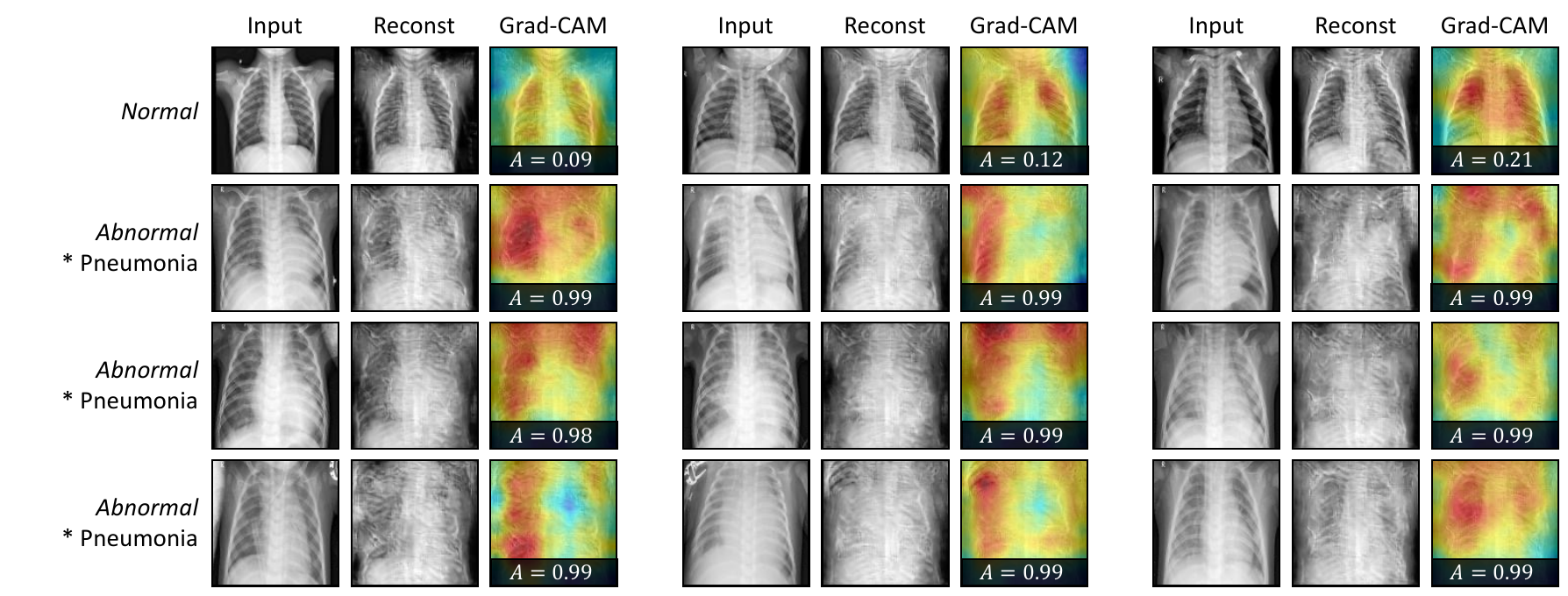}
    \caption{[Better viewed on-line, in color, and zoomed in for details]
    Reconstruction results of \methodname\ on the ZhangLab dataset. The corresponding Grad-CAM heatmaps, along with anomaly scores, are shown. The anomaly score denotes the probability of the image containing abnormal.
    }
    \label{fig:recon_results_chexpert}
\end{figure*}

\subsection{Benchmarking \methodname\ on Three Public Datasets}
\label{sec:benchmark_chest_xray}

Our \methodname\ was mainly evaluated on three large-scale benchmarks: ZhangLab Chest X-ray, COVIDx and Stanford CheXpert for comparing with a wide range of state-of-the-art methods. According to \tableautorefname~\ref{tab:chest_xray_benchmark}, our \methodname\ achieves the most promising result in terms of most metrics on these datasets. Specifically, \methodname\ outperforms the second best runner-up counterpart SALAD~\cite{zhao2021anomaly} by 8.4\% in AUC, 9.1\% in Accuracy, and 5.9\% in F1 on the ZhangLab dataset. In particular, our \methodname\ trained in an unsupervised manner surpasses Fixed-Point GAN~\cite{siddiquee2019learning}---a weakly supervised anomaly detection method---by 8\% in AUC. 
Additionally, CutPaste~\cite{li2021cutpaste} and M-KD~\cite{salehi2021multiresolution} were previous state of the arts on COVIDx and CheXpert datasets, respectively.
Our \methodname\ not only achieves 5.0\% and 9.9\% improvements, but also significantly exceeds its previous version (SQUID).
The ROC curve and PR curve are presented in \figureautorefname~\ref{fig:roc}, demonstrating that our method yields the best trade-off between sensitivity and specificity. 
Overall, the significant improvements observed with \methodname\ proved the effectiveness of our proposed designs and techniques in this work. 

In \figureautorefname~\ref{fig:recon_results_chexpert}, we visualize the reconstructions of \methodname\ on exemplary normal and abnormal images in the ZhangLab dataset. For normal cases, \methodname\ can easily find a similar match in the memory and hence achieves the reconstruction smoothly. For abnormal cases, the contradiction will arise by imposing forged normal patterns into the abnormal features. In this way, the generated images will vary significantly from the input, which will then be captured by the discriminator. We plot the heatmap of the discriminator (using Grad-CAM~\cite{selvaraju2017grad}) to indicate the regions that are poorly reconstructed. As a result, the reconstructed healthy images yield much lower anomaly scores than the diseased ones, validating the effectiveness of \methodname.

We also benchmarked the running speed of models to compare the efficiency of SQUID and the proposed SimSID. It is observed that one step training of SimSID is 17.2s faster than SQUID (11.0s \emph{v.s.} 28.2s) and one step inference of SimSID is 0.5s faster than SQUID (9.0s \emph{v.s.} 9.5s).

\noindent
\textbf{Limitation:} We found \methodname\, in its current form, is not able to \emph{localize} anomalies at the pixel level precisely. It is understandable because, unlike \cite{wang2017chestx,singh2017hide,zhang2018adversarial,tang2018attention,salehi2021multiresolution}, our \methodname\ is an unsupervised method, requiring zero manual annotation for normal/abnormal images. More investigation on pixel-level localization (or even segmentation) and multi-scale detections could be meaningful in the future.

\begin{table}[!t]
\footnotesize
\centering
    \caption{
    Component studies indicate that the overall performance benefits from all the components in \methodname. The ablation study is conducted on the Zhanglab dataset.} 
    \begin{tabular}{p{0.42\linewidth}P{0.13\linewidth}P{0.13\linewidth}P{0.13\linewidth}}
        Method & AUC(\%) & Acc(\%) & F1(\%) \\
        \shline
        \emph{w/o} Space-aware Memory & 77.6$\pm$0.5& 75.5$\pm$0.5 & 82.5$\pm$0.6 \\
        \emph{w/o} Skip Connection & 79.5$\pm$1.6 & 73.0$\pm$1.4 & 78.8$\pm$0.5 \\
        \emph{w/o} In-painting Block & 80.9$\pm$2.1 & 75.8$\pm$1.5 & 81.6$\pm$1.3\\
        \emph{w/o} Hierarchical Memory & 82.9$\pm$1.2 & 77.4$\pm$1.1 & 81.2$\pm$0.5 \\
        \emph{w/o} Knowledge Distillation & 85.4$\pm$0.8 & 79.5$\pm$0.7 & 83.5$\pm$0.8 \\
        \emph{w/o} Stop Gradient & 85.0$\pm$4.3 & 77.6$\pm$2.8 & 79.8$\pm$1.6 \\
        \emph{w/o} Gumbel Shrinkage & 86.2$\pm$3.3 & 80.5$\pm$3.2 & 85.4$\pm$2.1 \\
        \emph{w/} Memory Queue instead & $86.7\pm2.1$ & $80.6\pm1.7$ & $84.2\pm1.3$\\ 
        \hline
        Convolution Layers & 86.3$\pm$3.4 & 80.8$\pm$3.0 & 85.4$\pm$2.2 \\
        Pixel-level In-painting & 79.1$\pm$0.4 & 74.4$\pm$1.6 & 81.3$\pm$0.9 \\
        SQUID \cite{xiang2023squid} & 87.6$\pm$1.5 & 80.3$\pm$1.3 & 84.7$\pm$0.8 \\
        \hline
        Full \methodname & \textbf{91.1$\pm$0.9} & \textbf{85.0$\pm$1.0} & \textbf{88.0$\pm$1.1}\\
    \end{tabular}
    \label{tab:component}
\end{table}

\subsection{Ablating Key Properties in \methodname} 
\label{sec:ablation}

\medskip\noindent\textbf{Component study:} 
We first examine the impact of components in \methodname\ by taking each one of them out of the entire framework. 
\tableautorefname~\ref{tab:component} shows that each component accounts for at least 5\% performance gains.
The space-aware memory ($+13.5$\%) and in-painting block ($+10.2$\%) are among the most significant contributors, which underline our motivation and justification of the method development (\S\ref{sec:queue} and \S\ref{sec:inpaint}).
Moreover, the knowledge distillation from teacher to student generators strikes an important balance:
the student generator reconstructs faithful ``normal'' images from similar anatomical patterns in the dictionary while preserving the unique characteristics of each input image (regularized by the teacher generator).
Besides, we must acknowledge that the training tricks (\eg hard shrinkage~\cite{jang2016categorical}, stop gradient~\cite{he2020momentum}) are necessary for the remarkable performance.

\begin{figure}[!t]
    \centering
    \includegraphics[width=1.0\linewidth]{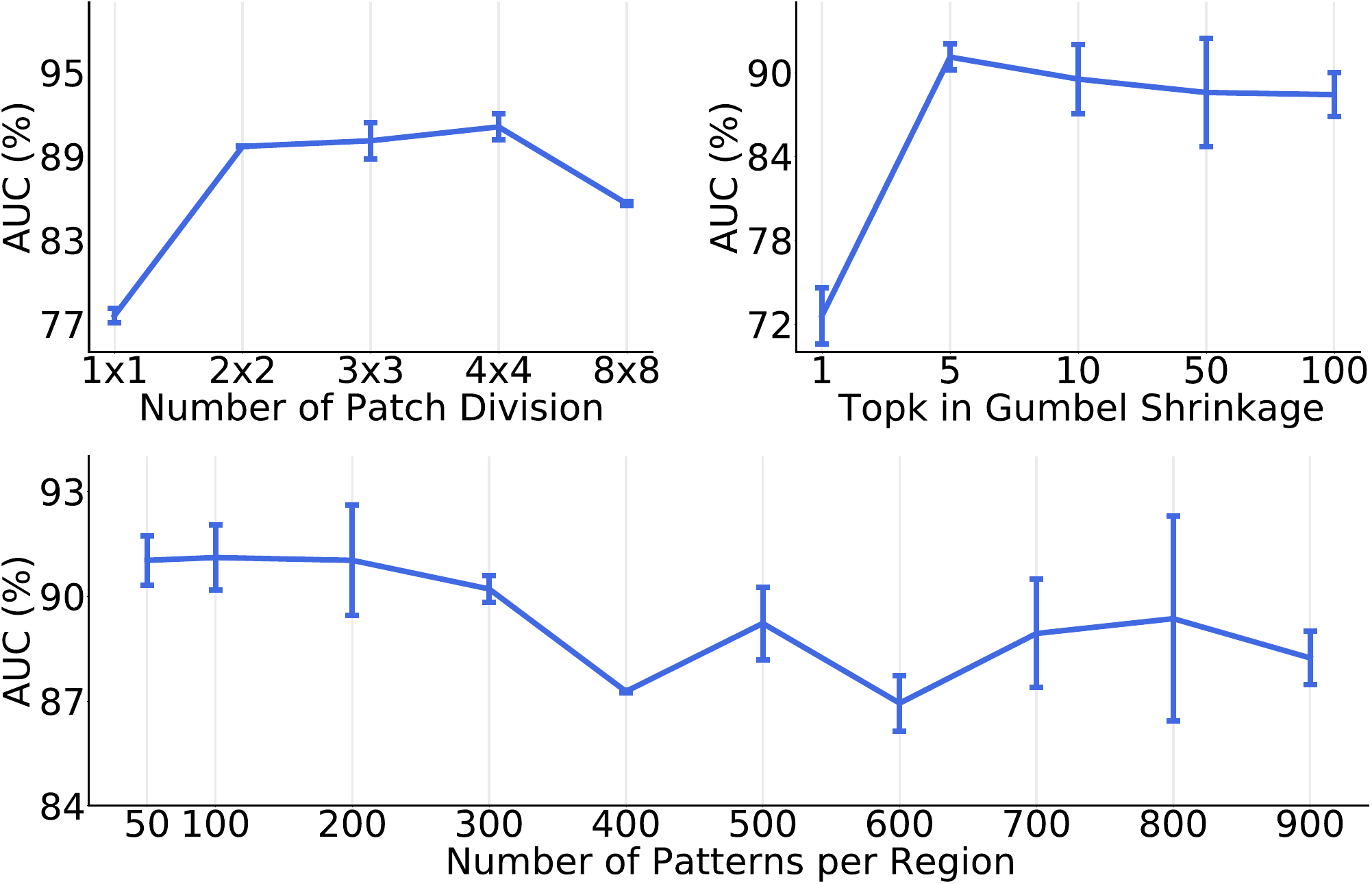}
    \caption{
    \methodname\ is robust to hyper-parameter modifications to some extent. The best result is obtained by dividing $4\times4$ patches, setting 100 patterns per memory region, and activating the top 5 patterns through Gumbel Shrinkage. The hyper-parameters were determined on the validation set of ZhangLab and were applied to all three datasets.
    }
    \label{fig:ablation}
\end{figure}

We further ablate the feature in-painting design in our model by comparing to other reasonable module designs:
In our proposed in-painting block, a transformer layer is used to aggregate the patch features and the Memory Matrix augmented ``normal'' features. However, one may wonder if a simple convolution layer can also suffice. We conducted experiments by replacing the transformer layer with a convolution layer while preserving other structures. The result is presented in the 8th line of \tableautorefname~\ref{tab:component}, where with convolution layer the AUC decreased by 4.8\%.
Another comparable design is pixel-level in-painting. As discussed in~\S\ref{sec:inpaint}, raw images usually contain larger noise and artifacts than features, so we proposed to achieve the in-painting at the feature level rather than at the pixel level~\cite{li2020recurrent,pathak2016context,zhou2021models,pinaya2022unsupervised}. To validate our claim, we have conducted experiments on carrying out the in-painting at the pixel level. Instead of using a transformer layer to in-paint the extracted patch features, we randomly zeroed out parts of the input patches with 25\% probability and let \methodname\ in-paint the distorted input images. All other settings and objective functions remain unchanged. The result is shown in the 10th row of \tableautorefname~\ref{tab:component}, and feature-level in-painting surpasses pixel-level in-painting by 12.0\% in AUC. As shown from the table, we validate that the new space-aware memory matrix, the in-painting block, the hierarchical memory design and skip connections are among the most important contributions to performance. Based on the proposed block, all other components can be combined in a more effective way than SQUID. We attribute the improvements to better feature representation. With the memory design in this paper, the memory matrix is learned together with other model components, and learns a condensed feature representation for the whole training set. While with the original memory queue design in SQUID, the limited-size queue is only able to record a fraction of features in the training set, and these features are biased toward specific samples. Therefore, the improvement over SQUID comes from the memory features that better encode feature patterns in the training set.

\medskip\noindent\textbf{Hyper-parameter robustness:} The number of patch divisions, the topk value in Gumbel Shrinkage, and the number of memory patterns within a specific region of Memory Matrix are three important hyper-parameters of \methodname. Here, we conducted exhausted experiments on these parameters in \figureautorefname~\ref{fig:ablation}. Trials were first made on the number of patches from $1\times1$ to $8\times8$. When dividing input images into a single patch, space-aware settings are not triggered, hence yielding the worst performance. Although the spatial structures are relatively stable in most chest X-rays, certain deviations can still be observed. Therefore, with small patches, object parts in one patch can easily appear in adjacent patches and be misdetected as anomalies. Note that the best setting of patch number differs from SQUID, which attributes to the modification of the in-painting block. We found that Memory Matrix is less robust to spatial difference than Memory Queue. When segmenting an input image into non-overlapping patches, more patch segments lead to a smaller size for each patch, and, eventually, less inconsistency per patch. Therefore, the modified memory matrix in the in-painting block of SimSID benefits from more patches. The number of topk activations in Gumbel softmax also impacts the performances. By assembling the top-5 most similar patterns through Gumbel softmax, \methodname\ is able to achieve the best result. When replacing input features with the top-1 most similar pattern, \methodname\ suffers from a performance drop by -18.5\% AUC. According to the AUC vs. number of patterns in each Memory Matrix region, we found that a small number of items is sufficient to support normal pattern querying in local regions and the best result is achieved by using merely 100 items per region. Degraded performance is observed at a greater number of items per region ($>$500).

\begin{figure}[!t]
    \centering
    \includegraphics[width=1.0\linewidth]{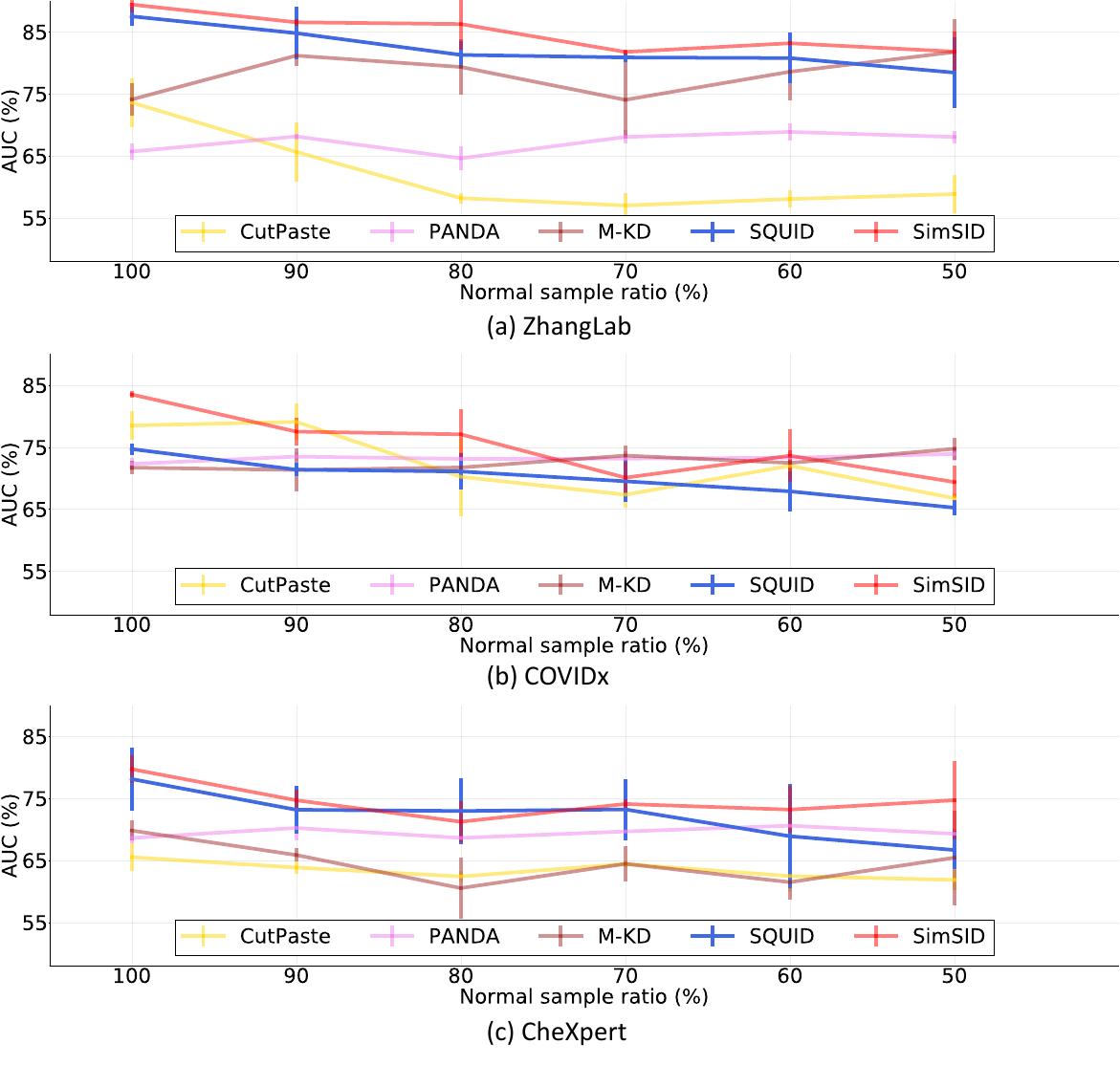}
    \caption{Ablation study of mixing normal and abnormal samples in the training set. \methodname\ is robust to mixed training with different normal / abnormal ratios on the ZhangLab, COVIDx and CheXpert datasets.
    }
    \label{fig:ratio_robustness}
\end{figure}

\subsection{Robustness to Abnormal Data in the Training Set}
\label{sec:impact_ratio}

Strictly speaking, existing unsupervised anomaly detection methods (\eg \cite{baur2021autoencoders,tian2022unsupervised}) are not unsupervised because they require a training set to be all ``normal''. To form this normal training set, image-level annotation as weak supervision is an implicit requirement. To the best of our knowledge, there is no work investigating the robustness of such ``unsupervised'' anomaly detection methods to the normal/abnormal ratio in real-world datasets.
With disease-free sample ratio in the training set ranging from 100\% to 50\%, we have compared the robustness of \methodname\ with four competitive baselines (SQUID~\cite{xiang2023squid}, CutPaste~\cite{li2021cutpaste}, PANDA~\cite{reiss2021panda} and M-KD~\cite{salehi2021multiresolution}) that originally relies on a pure normal training set.

\figureautorefname~\ref{fig:ratio_robustness} remarks that our proposed method is robust to the abnormal/normal training ratio up to 50\% and remains AUC above 0.8 on the ZhangLab dataset by automatically omitting minority anatomical patterns. On the Stanford CheXpert and the COVIDx datasets, \methodname\ still achieves better or comparable AUC to abnormal training samples than the baseline models.
We ask: \textit{When the training set does not contain exclusively normal images, how does \methodname\ discriminate between abnormal and normal patches?}
As described in \S\ref{sec:method}, we divide an image into small patches and the model predicts every patch feature based on its eight surrounding patches.
If a neighbor patch is abnormal, the other neighbors will contribute more to the in-painting process. 
Besides, abnormalities are often small, so the abnormal patches only account for a small proportion of an image, not to mention within the entire dataset.
Since most cropped patches are normal, the abnormal patches would not have a serious effect, as evidenced by our robust detection results up to a 50\% normal ratio.
In contrast, CutPaste drops significantly as the percentage of disease-free images decreases; PANDA and M-KD can maintain their performance due to the use of pre-trained features. Interestingly, M-KD with mixed data even outperforms its vanilla training setting, although with considerable fluctuations. SQUID, on the other hand, benefits from the neighborhood in-painting design, but is still consistently worse than SimSID. 

\section{Conclusion}

We present \methodname\ for unsupervised anomaly detection from radiography images.
The assumption behind our design is that radiography imaging protocols focus on particular body regions, therefore producing images of great similarity and yielding recurrent anatomical structures across patients. 
\methodname\ exploits the structural consistency of chest anatomy with the help of \textit{space-aware memory matrix} and \textit{feature in-painting}.
Qualitatively, we show that \methodname\ can taxonomize the ingrained anatomical structures into recurrent patterns; and in the inference, \methodname\ can identify anomalies (unseen/modified patterns) in the image.
Quantitatively, \methodname\ surpasses the state of the arts in unsupervised anomaly detection by +8.0\%, +5.0\%, and +9.9\% AUC scores on ZhangLab, COVIDx, and CheXpert benchmark datasets, respectively.

\section*{Acknowledgments}
This work was supported by the Lustgarten Foundation for Pancreatic Cancer Research and the Patrick J. McGovern Foundation Award. We thank Xiaoxi Chen for annotating the Chest X-ray in \figureautorefname~\ref{fig:introductory_figure}.

\bibliographystyle{IEEEtran}
\bibliography{refs,zzhou}

\end{document}